\documentclass[prl,twocolumn,showpacs,nofootinbib]{revtex4}
\usepackage{epsfig,amsmath,bm}
\begin{document}

\title{Aspects of the confinement mechanism in Coulomb-gauge QCD}

\author{R.~Alkofer}
\affiliation{Institut f\"ur Physik, University of Graz, A-8010 Graz, Austria}

\author{M.~Kloker}
\affiliation{Institut f\"ur Physik, University of Graz, A-8010 Graz, Austria}

\author{A.~Krassnigg}
\affiliation{Institut f\"ur Physik, University of Graz, A-8010 Graz, Austria}

\author{R.F.~Wagenbrunn}
\affiliation{Institut f\"ur Physik, University of Graz, A-8010 Graz, Austria}

\date{\today}

\begin{abstract}
Phenomenological consequences of the infrared singular, instantaneous part of the gluon propagator
in Coulomb gauge are investigated. The corresponding quark Dyson-Schwinger equation is solved, neglecting retardation
and transverse gluons and regulating the resulting infrared singularities. While the quark propagator
vanishes as the infrared regulator goes to zero, the frequency integral over the quark propagator stays
finite and well-defined. Solutions of the homogeneous Bethe-Salpeter equation for the pseudoscalar and 
vector mesons as well as for scalar and axial-vector diquarks are obtained. In the limit of a vanishing 
infrared regulator the diquark masses diverge, while meson properties and diquark radii remain finite
and well-defined. These features are interpreted with respect to the resulting aspects of confinement
for colored quark-quark correlations.
\end{abstract}


\maketitle

The substructure of the nucleon has been determined to an enormous precision leaving no doubt that the parton picture
emerges from quarks and gluons, the elementary fields of Quantum Chromodynamics (QCD). Although they are the
``elementary particles'' of strong interactions, quarks and gluons have never been detected outside hadrons. This
phenomenon is called confinement. Despite its importance for particle physics and for an axiomatic approach to
quantum field theory our understanding of confinement is far from being satisfactory.

In this letter we concentrate on certain aspects of confinement for colored composite states. We start from
the commonly accepted Wilson criterion \cite{Wilson:1974sk} and an inequality between the gauge-invariant
quark-antiquark potential $V_W(R)$ and the color-Coulomb potential $V_C(\vec{x})$ 
\cite{Zwanziger:2002sh}. The latter quantity is the instantaneous part of the time-time component of the gluon
propagator in Coulomb gauge: $D_{00}(\vec{x},t)\propto V_C(\vec{x})\,\delta(t)+$ non-inst.~terms.
In Ref.~\cite{Zwanziger:2002sh} it was shown that if $V_W(R)$ is confining, i.\,e.~if 
$\lim_{R\rightarrow\infty}V_W(R)\rightarrow\infty$, then also $|V_C(\vec{x})|$ is confining.
This was confirmed in an $SU(2)$ lattice calculation \cite{Greensite:2004ke} where it was found that 
$-V_C(\vec{x})$ rises linearly with $R=|\vec{x}|$. However, the corresponding string tension, $\sigma_c$,
was extracted to be several times the asymptotic one\footnote{If the same holds for the physical case of three colors
one infers $\sigma_c\approx 600\ldots 750$ MeV from the generally used value $\sigma_c\approx 440$ MeV. Note, however,
that this increase is not sufficient to resolve the problem of a too small value of the pion decay constant
\cite{Adler:1984ri}, when only a confining potential is used and non-instantaneous interactions, in particular
transverse gluons, are neglected.}. 

A well-suited formalism for the study of composite or bound states of quarks is the Dyson-Schwinger/Bethe-Salpeter
approach \cite{Roberts:2000aa
}. While corresponding investigations in Coulomb gauge, 
e.\,g.~\cite{Finger:1981gm,
leyaouanc:1988}, predate those based on model studies employing Landau-gauge
QCD Green functions, the latter have been much more numerous and the corresponding studies explore a large number of
hadron observables, see e.\,g.~Refs.~\cite{Maris:1997tm,nucleon,Maris:1999ta,nucleonff,Maris:2002yu,Alkofer:2002bp} 
and references therein. Note that in Landau-gauge QCD the structure of the quark-gluon vertex 
\cite{Skullerud:2002ge
}
is an issue of current debate due to its importance for the quark propagator
\cite{Alkofer:2003jj
}.

In this letter we report on a study of mesons and two-quark composite states employing the color-Coulomb potential
$V_C(\vec{x})$ and thus some of the basic features of Coulomb-gauge QCD. We build on investigations
of the gluon propagator 
\cite{Szczepaniak:2003ve
}
and the dynamical breaking of chiral symmetry
\cite{Adler:1984ri,Finger:1981gm,
Llanes-Estrada:2004wr}
in Green-function approaches and related results of lattice calculations 
\cite{Greensite:2004ke,Cucchieri:2000gu
}. 
Our focus is the realization of confinement for quarks and two-quark composite states (``diquarks'').

First, we briefly review the quark Dyson-Schwinger (gap) and bound-state Bethe-Salpeter equations. 
All calculations are performed in Minkowski space. The QCD gap equation determines 
the quark self-energy due to gluons. It is of the form
\begin{equation}\label{gapequ}
i\;S^{-1}(p) = /\!\!\!p-m - \Sigma(p)\;,
\end{equation}
where $S(p)$ is the renormalized dressed quark propagator, $m$ the current-quark mass, and $\Sigma(p)$ is
the quark self energy. A quark-antiquark bound state
is described by the Bethe-Salpeter equation (BSE), which in its homogeneous form is written as
(for simplicity we neglect Dirac, flavor, and color indices)
\begin{equation}\label{bse}
\Gamma(P,q) = \int d^4k\;K(q,k,P)\;S(k_+)\;\Gamma(P,k)\;S(k_-)\;,
\end{equation}
where $P$ and $q$ are the quark-antiquark pair's total and relative four-momenta, $\Gamma(P,q)$ is the bound state's
Bethe-Salpeter amplitude (BSA), $k_\pm = k\pm P/2$ are the individual quark- and antiquark-momenta, and
$K(q,k,P)$ is the quark-antiquark scattering kernel. Note that the result of Eq.~(\ref{gapequ}) appears as input
in Eq.~(\ref{bse}).

\begin{figure}
\epsfig{file=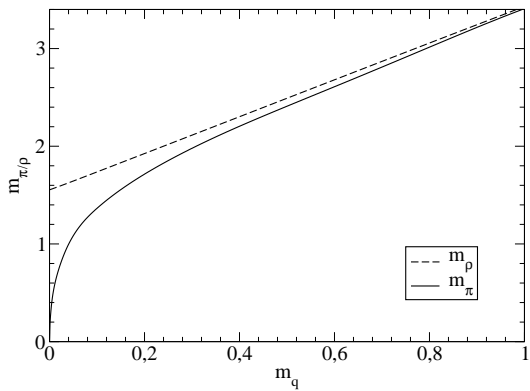,width=5.2cm,clip=true,angle=270}
\caption{The pion and rho masses as functions of the current-quark mass. All quantities are given in appropriate 
units of $\sqrt{\sigma_c}$, $\sigma_c$ being the Coulomb string tension (see text).\label{pionmass}}
\end{figure}
The quark self-energy in Eq.~(\ref{gapequ}) is a functional of the quark and gluon propagators and the quark-gluon
vertex; a self-consistent solution would require to simultaneously solve the Dyson-Schwinger equations for these
functions and the quark-antiquark scattering kernel from Eq.~(\ref{bse}) as well.
However, these equations again involve higher Green functions and therefore a truncation of this infinite coupled
system of integral equations is necessary.

In the present study we use Coulomb gauge together with an instantaneous
approximation and neglect the effects of transverse gluons. These approximations simplify the technical challenges
involved in concrete calculations. On the other hand, some component of the physics contained in the system is lost.
The results are qualitatively, but not quantitatively significant. We
therefore refrain from using physical dimensions, but instead present the quantities in all graphs 
in appropriate units of the Coulomb string tension $\sigma_c$. The reason for the qualitative
reliability of the calculations is that the underlying symmetries of the theory are incorporated
in the model via Slavnov-Taylor or Ward-Takahashi identities. One important example is the
axial-vector Ward-Takahashi identity, which is used to ensure that the kernels of the gap and
Bethe-Salpeter equations for pseudoscalar states are related in such a way that chiral symmetry and its
dynamical breaking are respected by the truncation.
In particular this leads to the correct behavior of the pion mass as a
function of the current-quark mass in the chiral limit. This behavior is shown in Fig.~\ref{pionmass}.
(Note: In the following we will in general present results for the chiral limit. The results for finite
current-quark mass are analogous.)
In this way one can reliably make qualitative statements about hadrons and their properties; however,
it is still important to investigate the contributions from retardation effects and transverse gluons,
and such efforts are currently made.

\begin{figure}
\epsfig{file=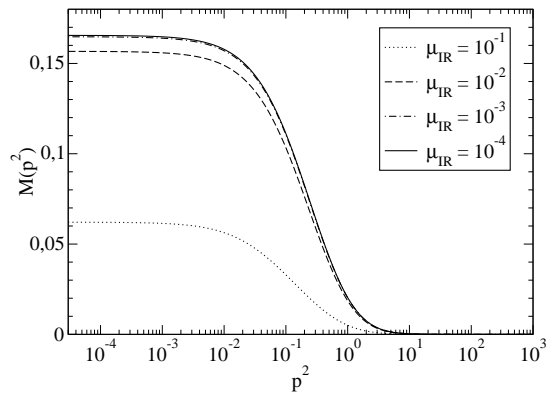,width=5.2cm,clip=true,angle=270}
\caption{The quark mass function $M(q^2)$ for four values of the infrared regulator $\mu_\mathrm{IR}$. All 
quantities are given in appropriate units of $\sqrt{\sigma_c}$.\label{massfplot}}
\end{figure}
In our model the quark self energy $\Sigma(p)$ in Eq.~(\ref{gapequ}) takes the form
\begin{equation}\label{qselfedetail}
\Sigma(p)=C_f\,6\pi\,\int\!\frac{d^4q}{(2\pi)^4}V_C(\vec{k})\,\gamma_0\,S(q)\,\gamma_0\;,
\end{equation}
where $C_f=(N_c^2-1)/(2 N_c)=4/3$ and $\vec{k}=\vec{p}-\vec{q}$.
In the following we will use  $p$ to denote $p=|\vec{p}|$.
The $q_0$-integration in Eq.~(\ref{qselfedetail}) can be performed easily. One makes the Ansatz 
$S^{-1}(p):=-i(\gamma_0p_0-\vec{\gamma}\cdot\vec{p}\;C(p) - B(p))$
and obtains two coupled integral equations for the functions $B(p)$ and $C(p)$
\begin{eqnarray}\label{gapequdetaila}
B(p)&=&m+\frac{1}{2\pi^2}\int d^3q\;V_C(k)
\frac{M(q)}{\tilde{\omega}(q)}\\\label{gapequdetailb}
C(p)&=&1+\frac{1}{2\pi^2}\int d^3q\;V_C(k)\;\hat{p}\cdot\hat{q}\;
\frac{q}{p\,\tilde{\omega}(q)}\;,
\end{eqnarray}
where $\hat{p}=\vec{p}/p$, $m$ is the 
current-quark mass, $\tilde{\omega}(p):=\sqrt{M^2(p)+p^2}$, and 
$M(q):=B(q)/C(q)$ is the quark ``mass function''.
Its infrared behavior is a result of dynamical chiral symmetry breaking and can be used to define a 
constituent-quark mass; we have plotted the mass function $M$ as a function of $q^2$ in 
Fig.~\ref{massfplot} (details of this figure will be specified below).

\begin{figure}
\epsfig{file=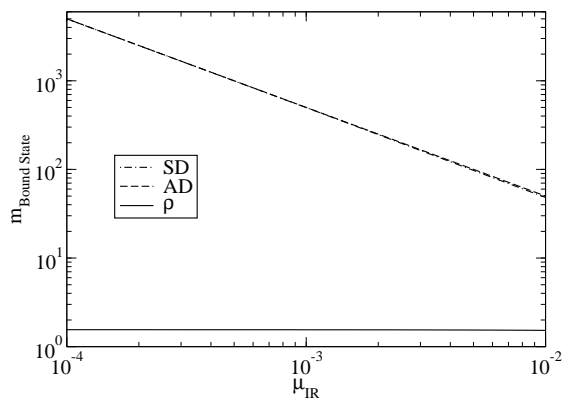,width=5.2cm,clip=true,angle=270}
\caption{The masses of the $\rho$ as well as the scalar (SD) and axial-vector (AD) diquarks as 
functions of the infrared regulator $\mu_\mathrm{IR}$. The mass of the $\pi$ is identically zero for all
values of $\mu_\mathrm{IR}$ and therefore not shown in the graph. All quantities are given in appropriate units of 
$\sqrt{\sigma_c}$.\label{diquarkmass}}
\end{figure}
The same approximations and conventions are used in the BSE. For pseudoscalar mesons in our model (and correspondingly scalar
diquarks) the BSA can be characterized in terms of two scalar functions $h(p)$
and $g(p)$, which essentially are the coefficients of the pseudoscalar and axial-vector structures
in the BSA. For details, see Ref.~\cite{Langfeld:1989en}.
The BSE, Eq.~(\ref{bse}), in terms of $h(p)$ and $g(p)$ in our model becomes
\begin{eqnarray}\label{bsedetail}
h(p)\,\omega(p)&=&\frac{1}{2\pi^2}\int\! d^3q\;V_C(k)\;\left[h(q)+\frac{m_\pi^2}{4\,\omega(q)}g(q)\right]\\
\nonumber
g(p)\,[\omega(p)\!\!\!&-&\!\!\!\frac{m_\pi^2}{4\,\omega(p)}]=h(p)+\\
\frac{1}{2\pi^2}\!\!\!&{\displaystyle\int}&\!\!\!\! d^3q\;V_C(k)\left[
\frac{M(p)\,M(q)+\vec{p}\cdot\vec{q}}{\tilde{\omega}(p)\,\tilde{\omega}(q)}\right]g(q)\;,
\end{eqnarray}
where $m_\pi$ is the bound state's (e.\,g.~the pion's) yet unknown mass and $\omega(p)=C(p)\,\tilde{\omega}(p)$.

For vector mesons (and correspondingly axial-vector diquarks)
the BSA has four linearly independent amplitudes. The construction of the four coupled 
integral equations corresponding to the BSE is analogous to the pseudoscalar case.

The Coulomb-gluon part $V_C$ of the interaction in Eqs.~(\ref{gapequdetaila}), (\ref{gapequdetailb}) 
and (\ref{bsedetail}) 
is chosen to be 
\begin{equation}\label{pot}
V_C(k) = \frac{\sigma_c}{(k^2)^2}\;,
\end{equation}
where $\sigma_c$ is the Coulomb string tension. 
Obviously, $V_C(k)$ is infrared singular. It is regulated by a parameter $\mu_\mathrm{IR}$ such that the  
momentum dependence is modified to
\begin{equation}\label{mupot}
V_C(k) = \frac{\sigma_c}{(k^2)^2}\rightarrow \frac{\sigma_c}{(k^2+\mu_\mathrm{IR}^2)^2}\;.
\end{equation}
\begin{figure}
\epsfig{file=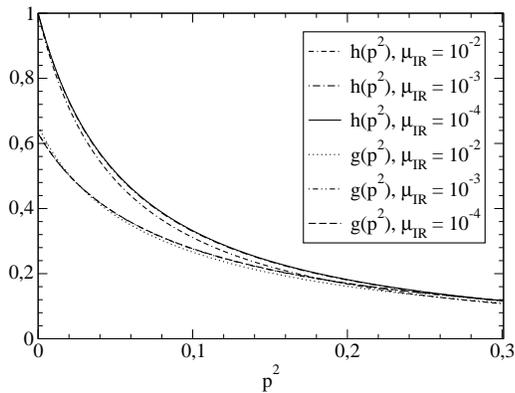,width=5.2cm,clip=true,angle=270}
\caption{Pion Bethe-Salpeter amplitude components $g$ and $h$ as functions of the infrared regulator $\mu_\mathrm{IR}$. 
For convenience, the amplitudes are normalized such that $h(0)=1$.
All quantities are given in appropriate units of $\sqrt{\sigma_c}$.\label{pibsaplots}}
\end{figure}
In this fashion all quantities and observables become $\mu_\mathrm{IR}$-dependent and one obtains the final result for
some $f(\mu_\mathrm{IR})$ by taking the limit $f=\lim_{\mu_\mathrm{IR}\rightarrow 0}f(\mu_\mathrm{IR})$. 
This is illustrated for the quark mass function in
Fig.~\ref{massfplot}: $M(p^2)$ is plotted for different values of $\mu_\mathrm{IR}$ and it is clear 
that the curves converge onto a final result for $\mu_\mathrm{IR}\rightarrow 0$.

In order to check the UV behavior one can use a Richardson potential \cite{Richardson:1978bt}, 
which has the momentum
dependence $V_C(k) \sim 1/(k^2\;\ln(1+k^2/\Lambda^2))$.
The advantage of our choice for $V_C$ is that the angular integration required to solve 
Eqs.~(\ref{gapequdetaila}), (\ref{gapequdetailb}) can be performed analytically. We have checked that the 
qualitative results presented in this paper can be reproduced with the Richardson potential. 
Details of this approach and its UV renormalization will be published elsewhere.

\begin{figure}
\epsfig{file=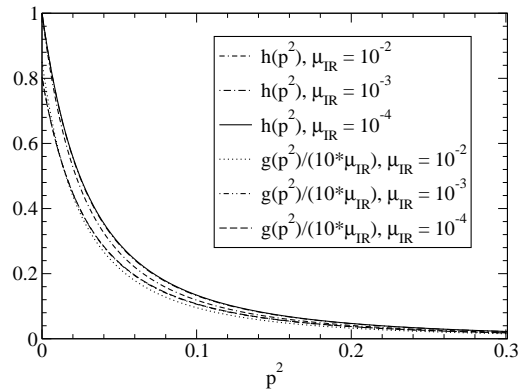,width=5.2cm,clip=true,angle=270}
\caption{Same as Fig.~\ref{pibsaplots} for the scalar diquark.\label{dibsaplots}}
\end{figure}
The homogeneous BSE in Eq.~(\ref{bse}) is solved by introducing an eigenvalue $\lambda(P^2=M^2)$ with
$M$ the bound-state mass. One then finds $M$ such that $\lambda=1$ (for mesons) and
$\lambda=2$ (for diquarks). For details, see e.\,g.~\cite{Maris:2002yu}.

The curve for $\lambda(P^2)$ gets less inclined with smaller values of the infrared regulator $\mu_\mathrm{IR}$, and
its intersection point with $\lambda=1$ stabilizes in the limit $\mu_\mathrm{IR}\rightarrow 0$. As a consequence,
while the meson mass is stable, the mass eigenvalue for the corresponding diquark state (corresponding to 
$\lambda(M)=2$) increases like $1/\mu_\mathrm{IR}$, ultimately completely removing these states from the 
physical spectrum.
We have illustrated these effects in Fig.~\ref{diquarkmass} for values of $10^{-4}\le\mu_\mathrm{IR}\le 10^{-2}$. 

Note: for $N_c=2$ diquarks correspond to baryons. In particular, in ladder approximation 
the respective color factors for meson and baryon BSEs are identical.
Therefore the properties of the scalar (axialvector)  baryon are identical to those of the pion ($\varrho$ meson). 
For $N_c\ge 3$ the ratio of quark-quark to quark-antiquark color
factors increases like $N_c-1$; this means that the argument given above is also valid in the large-$N_c$ limit.

\begin{figure}
\epsfig{file=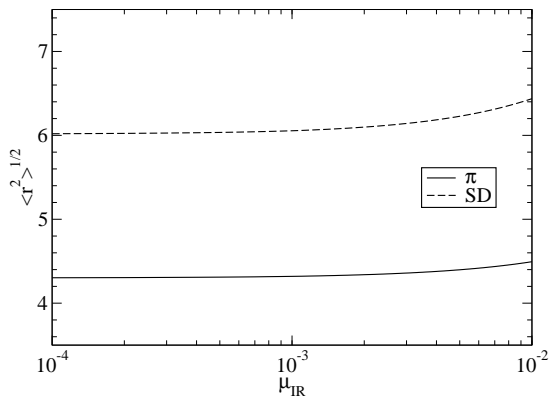,width=5.2cm,clip=true,angle=270}
\caption{Charge radii for $\pi$ meson as well as scalar diquark (SD) 
as functions of the infrared regulator $\mu_\mathrm{IR}$. All quantities are given in appropriate units of 
$\sqrt{\sigma_c}$.\label{chargeradii}}
\end{figure}
We studied the BSAs as $\mu_\mathrm{IR}\rightarrow 0$: the results for $g$ and $h$ (\ref{bsedetail}) 
are presented in Figs.~\ref{pibsaplots} and \ref{dibsaplots} for the pion and scalar diquark, 
respectively. For convenience, the normalization of the amplitudes has been chosen 
such that $h(0)=1$. We note, however, that IR-cancellations appearing in the pion case lead to a stable $h$ as
well as ratio
of $g/h$, which is not the case (as one would naively expect) in the diquark case: there 
$g/h\sim \mu_\mathrm{IR}\rightarrow 0$ and $h\sim 1/\sqrt{\mu_\mathrm{IR}}$. 
Still, one can investigate the charge radii 
for both meson and diquark states by requesting that the electromagnetic form factor at the origin
yields the bound-state charge, which gives finite results in the limit $\mu_\mathrm{IR}\rightarrow 0$. 
Plots of the pion and scalar diquark charge radii are shown in Fig.~\ref{chargeradii}.
The results for vector-meson and axial-vector-diquark amplitudes are analogous.

We have performed a study of pseudoscalar- and vector-meson states and their corresponding diquark
partners in a simple model of Coulomb-gauge QCD in the context of Dyson-Schwinger equations, which allows for 
obtaining reliable qualitative information about hadrons. 
The infrared singularities in the integrands are regulated by
the scale $\mu_\mathrm{IR}$ such that final results are obtained in the limit 
$\mu_\mathrm{IR}\rightarrow 0$. In this limit
the masses and charge radii for the mesons are stable; for their diquark partners only
the masses diverge like $1/\mu_\mathrm{IR}$, while the charge radii do not. Thus the diquarks are
removed from the physical spectrum reflecting confinement of colored quark-quark correlations.
Nevertheless they possess a well-defined size. This adds to the motivation of
nucleon studies in a covariant quark-diquark picture \cite{nucleon,nucleonff}.

\begin{acknowledgments}
We thank C.\,D. Roberts and D. Zwanziger for valuable discussions and the critical reading of the manuscript.
This work was supported by: Deutsche Forschungsgemeinschaft under contract Al 279/5-1,
a grant from the Ministry of Science, Research and the Arts of
Baden-W\"urttemberg (Az: 24-7532.23-19-18/1 and 24-7532.23-19-18/2), and the 
Austrian Science Fund FWF, Schr\"odinger-R\"uckkehrstipendium R50-N08.
\end{acknowledgments}

\end{document}